# RADIATIVE p$^{11}$B CAPTURE AT ASTROPHYSICAL ENERGIES

S. B. Dubovichenko[1]

[1] *Fessenkov Astrophysical Institute "NCSRT" NSA RK, 050020, Almaty, Kazakhstan*
*dubovichenko@gmail.com*

N.A. Burkova [2, *†]

[2] *Al-Farabi Kazakh National University 050000, Almaty, Kazakhstan*
*natali.burkova@gmail.com*

In the framework of the modified potential cluster model with forbidden states the possibility of describing the available experimental data on the total cross sections and astrophysical S-factor for $p^{11}$B radiative capture to the ground state of $^{12}$C was treated at astrophysical energies.

**Introduction**

We are following the statement that the possibilities of simple potential cluster models (PCM) are far from exhausted. Furthermore, if they are based on the concept of forbidden states (FS) in two-body potentials [1], accounting the low energy resonance behavior of elastic scattering phase shifts directly, and fit the basic characteristics of clusters binding states (BS) in nuclei [2, 3], then PCM may be recognized as modified potential cluster models (MPCM).

As it was shown in [2, 3] the construction of rather simple version of MPCM is available while using the classification of orbital states by Young schemes what leads naturally to the introduction of forbidden states and explanation of resonances in elastic scattering phase shifts. This very approach enables to reproduce quite adequate a lot of available experimental on the total cross sections of the thermonuclear processes like radiative capture.

Therefore, just continuing the study of radiative capture reactions [1-5] entering into various thermonuclear cycles [6] we are going to treat the process $p^{11}B \to \gamma^{12}C$ at astrophysical energies within MPCM. There are no phase shift analyses for the scattering channel, so the interaction potentials might be constructed basing on the correct description of the resonance structure of $^{12}$C presented as $p^{11}$B bi-cluster system.

For BS or ground states (GS) of synthesized final nuclei with the same structure as in initial channel the interaction potentials are fitted basing on the accurate reproducing of the corresponding binding energy and some other main characteristics of these states, for example, charge radius and asymptotic constants (AC) [2, 3].

The choice of cluster model for the examining of such systems in nuclei, nuclear, and thermonuclear reactions at astrophysical energies is conditioned by the fact that in many of light nuclei the probability of clustering is very high. This is confirmed by the numerous experimental measurements and various theoretical calculations done by different authors during last fifty-sixty years [7, 8].

It is clear that assumption on the pure cluster configuration of a nucleus defined as two-body system of isolated associations in BS is a rather ideal situation. Therefore, the success of the potential model depends on the degree of the clusterization of a nucleus $A$ in channel with $A_1 + A_2$ nucleons.

In case of the dominating cluster configuration with minor impurity of others one-channel cluster model allows to identify this channel and reproduce the main nuclear features [9-12]. Methods of construction of the corresponding two-body potentials with FS are presented in details in [3, 13].

Besides, while speaking about the probability of some specific cluster channel in the nucleus one bears in mind its some definite stationary time-independent state. At a time, the radiative capture reactions are the time-dependent processes. So, one may assume that initial scattering channel with definite interacting clusters practically fixes the final bounded configuration of the formed final nucleus with practically 100% probability. We regard this argument as the explanation of reasonable success of MPCM in description of the total cross sections of the radiative capture processes, i. e. more than 20 [2-5, 9-12].

**1. Structure of cluster states**

In view of absence of the complete tables for the product s of Young schemes for the systems with a number of nucleons more than eight, [14] the presented below results might be consider as the qualitative estimation of the possible orbital symmetries in the ground state of $^{12}$C in the single taken $p^{11}$B cluster channel.

At the same time, just basing on such classification we succeeded to reproduce, and, what is more important, to explain available experimental data on the radiative nucleon capture reactions in channels $p^{12}$C [2] and $p^{13}$C [15]. So, it is reasonable to apply our approach basing on the classification of cluster states by orbital symmetry what leads to appearing of a number of FS and allowed states (AS) in partial two-body potentials, and as a consequence relative motion wave functions (in present case those refer to the proton and $^{11}$B system) have a set of nodes.

Let us assume the {443} orbital Young scheme for $^{11}$B, then within the 1p-shell treating of $p^{11}$B system one has {1} × {443} → {543} + {444} + {4431} [14]. The first scheme in this product is compatible with the orbital momenta $L$ = 1, 2, 3, 4, it is forbidden, as no more than four nucleons may be on s-shell. Second scheme is allowed and compatible with the orbital momenta $L$ = 0, 2, 4, as for the third one also allowed the corresponding momenta are $L$ = 1, 2, 3 [16].

It should be noted that even such qualitative analysis of orbital symmetries allow to define that there are forbidden states in $P$- and $D$ – waves, and no ones in $S$ – state. Actually, such a structure of FS and AS allows the construction of two-body interaction potentials required for the calculations of total cross sections for the treating reaction.

Thereby, just restricting by the lowest partial waves with orbital angular moments $L$ = 0 and 1 one may state that for $p^{11}$B system ($^{11}$B in GS has quantum numbers $J^{\pi}$, $T$ = 3/2⁻, 1/2 [17]) $^{3}S_{1}$ potential (here the notation $^{(2S+1)}L_{J}$ is used) has the allowed state only, which may be not bound and lay in continuous spectrum, as far as FS it is absent. Each of $^{3}P$ waves has bound forbidden and allowed state. One of them corresponds to the $^{3}P_{0}$ GS of $^{12}$C with quantum numbers $J^{\pi}$, $T$ = $0^{+}$, 0 and binding energy -15.9572 MeV [17] in $p^{11}$B channel. Others triplet $^{3}P$ states have bound FS, but may also have AS in continuous spectrum. Besides, there is mixture by spins $S$ = 1 and 2 may also appear both in scattering and bound states in $p^{11}$B system.

Besides GS let us treat two more resonance states appearing in $p^{11}$B system at positive energies

1. First resonance state appears at 162 keV (l.s.) or 148.6(4) keV in c.m. and yas a width less than 5.3(2) keV in c.m., and quantum numbers $J^{\pi}$ = $2^{+}$ (Tables 12.11 and 12.6 in [17]). It corresponds to 16.1058(7) MeV level in $^{12}$C and may be compare to mixed in spin $^{3+5}P_{2}$ wave with FS.

2. Second resonance state appearing at 675 keV (l.s.) has the width 300 keV in c.m. and quantum numbers $J^\pi = 2^-$ [17]. It corresponds to the level with 16.576 MeV and might be associated with $^5S_2$ scattering wave without FS.

Next resonance is at 1.388 MeV, i. e. higher than 1 MeV, so it will not be considered. In spectrum of $^{12}$C below 1 MeV there are no resonance levels which may be correlated to $^3S$ or $^5S$ scattering resonances [17]. That is why the corresponding phase shifts may be taken close to zero. There are no FS in $S$ wave, so the potentials for both spin channels $S = 1$ and $S = 2$ may be regarded zero also [3]. Thus, the minimum set of electromagnetic transitions to the GS will be considered in order to reproduce the energy dependence of the measured total cross sections.

So far as the ground state of $^{12}$C is $^3P_0$ level it is actual to include in consideration dipole electric $E1$ transition from the none-resonating $^3S_1$ scattering wave with zero interaction potential to the GS:

1. $^3S_1 \to {}^3P_0$.

Besides, $E2$ transition from the triplet part of $^3P_2$ scattering wave to the GS also should be evaluated as it has the resonance at 162 keV:

2. $^3P_2 \to {}^3P_0$.

There are a lot of details on the methods of calculation of cross sections in the framework of MPCM [2, 3, 18-20], so we are dropping this part here. Same concerns the methods of construction of two-body interaction cluster potentials at the given orbital momentum $L$, so we refer to [2, 3, 21].

In present calculations the following values of masses have been used – $m_p$ = 1.00727646677 [22] and $m(^{11}B)$ = 11.0093052 a.m.u. For the constant $\hbar^2/m_0$ the numerical value 41.4686 MeV·Fm$^2$ was taken, and despite it seems to be outdated we continue use it as far as keeping the consistency in comparison of the last [2, 3, 9,10] and all early presented results [19, 20].

**2. Interaction potentials**

For all $p^{11}$B potentials the Gauss form was used

$$V(r) = -V_0 \exp(-\alpha r^2), \qquad (1)$$

Let us now give the corresponding parameters for the GS and two resonating states of $^{12}$C presented as $p^{11}$B system. The following parameters have been found for the resonating $^{3+5}P_2$ wave with FS and $J = 2^+$

$$V_0 = 24.38058 \text{ MeV}, \quad \alpha = 0.025 \text{ Fm}^{-2}. \qquad (2)$$

Within this potential the resonance energy level $E$ = 162.0 (1) keV in *l.s.* and its width 0.8(1) keV in *c.m.* according [17] (it should be noted that this ref. gives the proton width of this level as 0.0217(18) keV in *l.s.*) data have been reproduced exactly: at level energy phase shift turned to be 90.0°(1). To calculate the level width basing on the phase shift we used expression $\Gamma = 2(d\delta/dE)^{-1}$. Energy dependence of $^{3+5}P_2$ phase shift is given in Fig. 1.

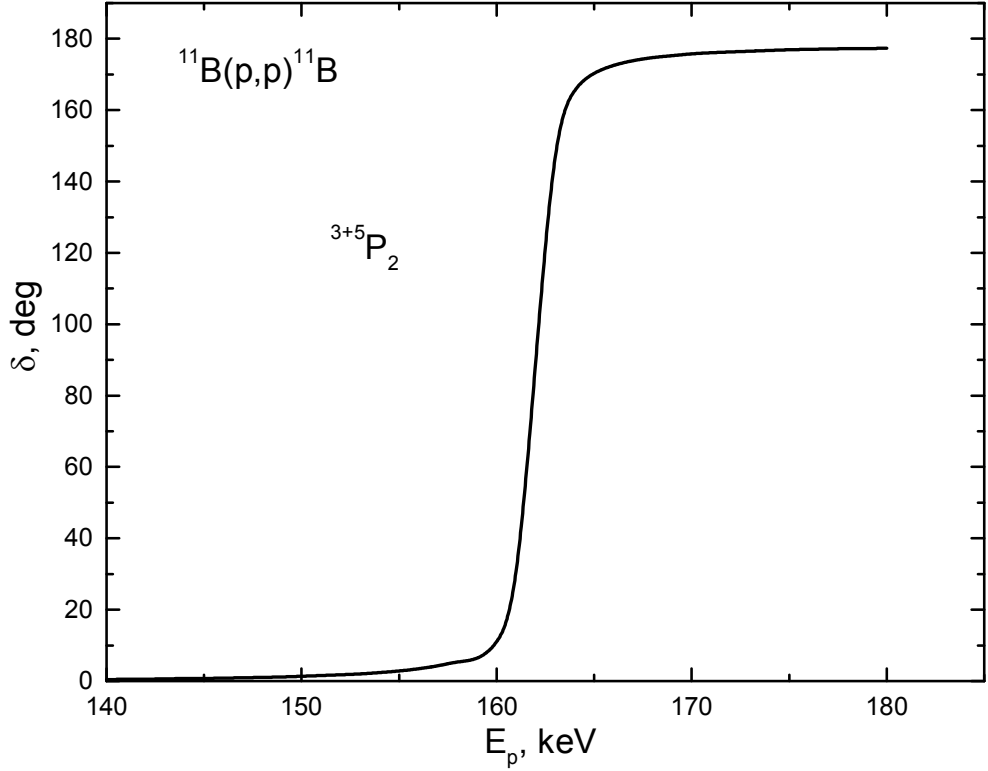

Fig. 1. Elastic $p^{11}$B scattering $^5P_2$ phase shift with resonance at 162 keV.

One should keep in mind, that if potential contains $N + M$ forbidden and allowed states, then it obeys to the generic Levinson theorem [1], and the corresponding phase shifts at zero energy should start from $\pi \cdot (N + M)$ or from 180° in present case as FS is bound, but AS is not.

However, in Fig. 1 for more conventional representation, the results for $^{3+5}P_2$ phase shifts are given from zero, but not from 180°. Let us note, that using model does not allow to separate $^3P$ and $^5P$ parts of such a potential, and that is why the $P_2$ potential with spin mixture has been obtained. Further down we are going to use the potential (2) both for $^3P_2$ and $^5P_2$ scattering states in the calculations of $M1$ or $EJ$ transitions.

Let us note, the analysis of resonance scattering bellow 1 MeV when the number of BS is given, the interaction potential may be defined completely unambiguously. At given number of BS the potential depth is fixed by the resonance energy of the level, and its width is defined by the resonance width. The parameters error usually not more than accuracy of the definition of the corresponding level and it is about 3 ÷ 5%. Same remark refers to the reconstruction procedure of the partial potentials using scattering phase shifts and information of the resonances in spectrum of final nucleus [3].

For the none-resonating $^3P_0$ and $^{3+5}P_1$ scattering waves with FS the following parameters are suggested

$V_0 = 60.0$ MeV, $\alpha = 0.1$ Fm$^{-2}$. (3)

This potential leads to the phase shifts close to 180(1)° in the energy range from zero up to 1.0 MeV and has one bound FS.

The interaction potential of the none-resonating process may be also performed unambiguously basing on the data for the scattering phase shifts and number of allowed and forbidden states in bound channels. The accuracy of the definition of the potential parameters depends on the accuracy of the corresponding phase shifts extracted from the scattering data and may be of 20 ÷ 30%. But such a potential has no ambiguities as its depth is fixed by the number of BS coming from the classification by Young schemes, and its width is defined by the phase shifts shape.

While constructing the none-resonating scattering potential basing on the data for the nuclear energy spectra it is much more difficult to estimate the accuracy of its parameters definition even when the number of BS is fixed. Usually such a potential should give the scattering shift close to zero bellow 1 MeV, or slowly go down at least [3].

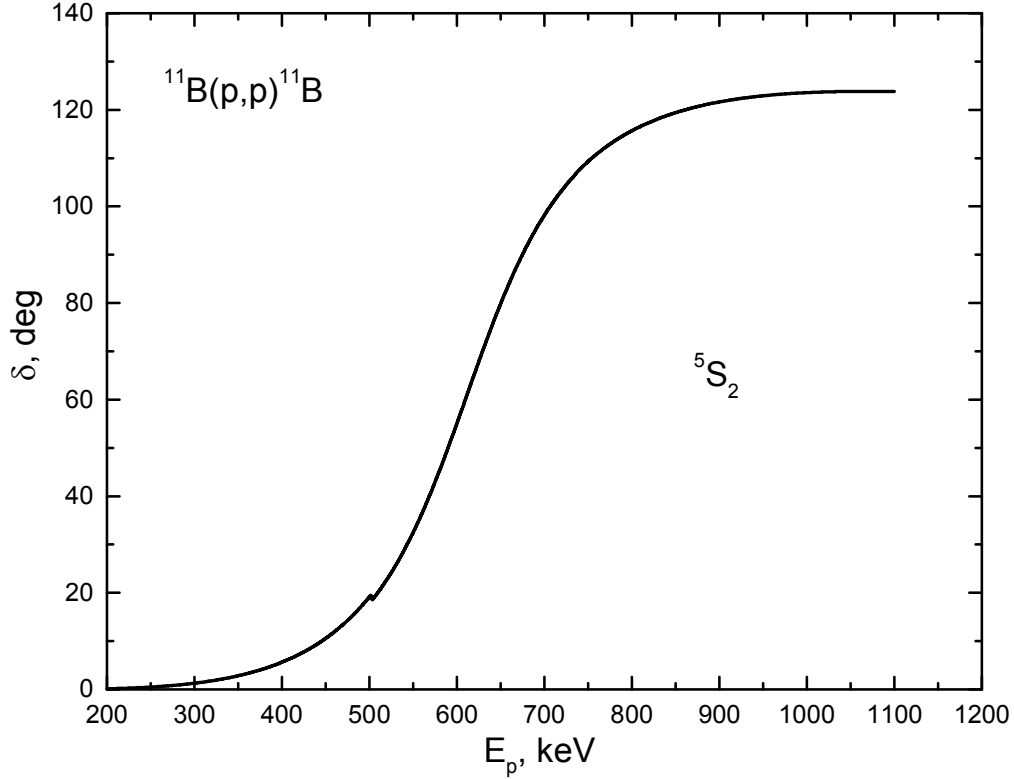

Fig. 2. . Elastic $p^{11}$B scattering $^5S_2$ phase shift with resonance at 675 keV.

For the resonating $^5S_2$ scattering wave without FS with $J = 2^-$ the following parameters are suggested

$V_0 = 10.9256$ MeV, $\alpha = 0.08$ Fm$^{-2}$, (4)

which lead to the phase shifts presented in Fig. 2. This potential reveals the resonance at 675(1) keV in *l.s.* with the width 289(1) keV I c.m. what is in agreement of data in [17], especially if keep in mind that given their value for the proton width is 150 keV in *l.s.*

For the potential corresponding to $^3P_0$ GS of $^{12}$C with FS in $p^{11}$B-cluster channel the following parameters were found

$$V_0 = 142.21387 \text{ MeV}, \alpha = 0.1 \text{ Fm}^{-2}. \tag{5}$$

This potential allows to obtain the mass radius $R_m$ = 2.51 Fm, charge radius $R_{ch}$ = 2.59 Fm, calculated binding energy -15.95720 MeV with an accuracy $\varepsilon = 10^{-5}$ MeV [24]. For the asymptotic constant (AC) written in the dimensionless form (see [25])

$$\chi_L(r) = \sqrt{2k_0} C_w W_{-\eta L+1/2}(2k_0 r)$$

value 23.9(2) was obtained over the range 7 – 13 Fm. Error of the calculated constant is defined by its averaging over the mentioned above distance range. For the mass and charge radii of $^{11}$B value 2.406(29) Fm [26] was used, radius of $^{12}$C was taken as 2.4702(22) Fm, charge and mass proton radius is 0.8775(51) Fm [22].

Scattering phase shift for such a potential does not exceed 0.1° at 1.0 MeV. As there is no any information on AC in $p^{11}$B channel in GS the potential (5) was constructed basing on the demands of reproducing of the none-resonating part of the total cross sections exclusively. It is clear that we should take into account the corresponding error bars, so in this very case the accuracy for the parameter definition pretends onto 10 % near. Let us emphasized, that in discussed case no independent data on AC have been found.

### 3. Total cross sections

First, dipole electric $E1$ transition $^3S_1 \to ^3P_0$ was treated when capture occurs from $^3S_1$ scattering wave with zero interaction central potential to the GS $^3P_0$ state constructed with potential (5). Calculated cross section is given by dash curve in Fig. 3 in the energy range 50 keV – 1.5 MeV. It fits well the non-resonant part of experimental data [27-30] kwon in the range 80 – 1500 keV.

Second, quadrupole electric $E2$ transition from the resonating at 162 keV $^3P_2$ wave to the GS was calculated. The sum of these two partial cross sections is shown by solid curve in Fig. 3, and well reproducing of the experimental data is obviously seen. At a time, it should be noted that calculated value of the cross section at the resonance energy 162 keV equals 101 $\mu b$ is essentially higher the measured one equals 5.5 $\mu b$. However, experimental data have been obtained at 163 keV, and as the resonance has very narrow width then the energy change even within one keV may lead to such an essential difference.

Note, that calculated cross section at 163 keV turned to be equal 20 $\mu b$ what is higher comparing the experimental value also. Meanwhile, the width of potential (2) 0.8 keV is considerably larger the proton width 0.02 keV from ref. [17]. Possibly this mismatch in resonance widths reveals in a rate of decreasing of the calculated cross sections.

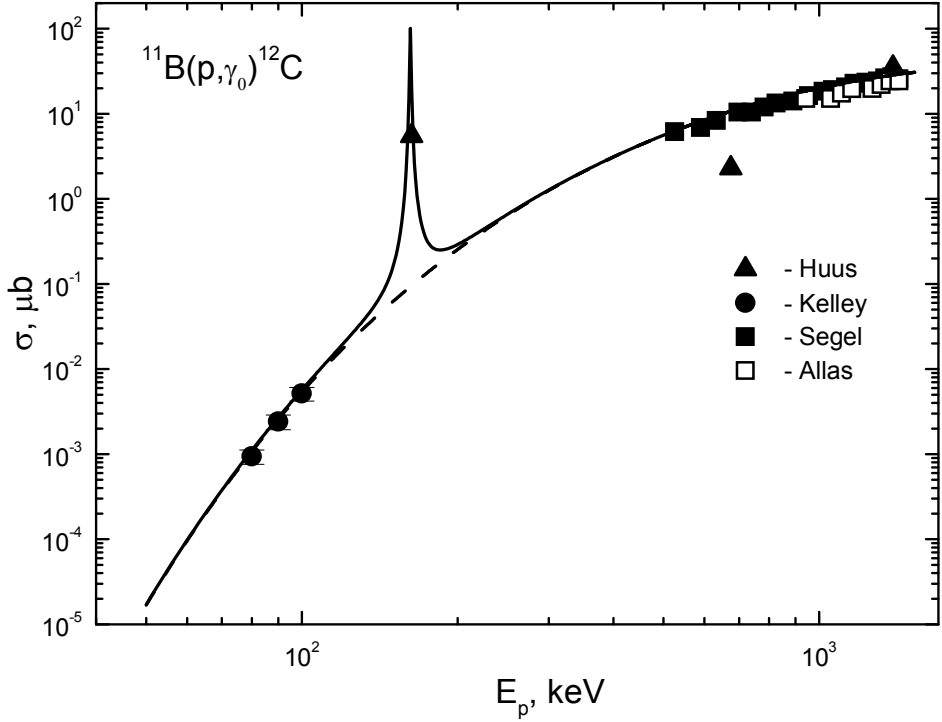

Fig. 3. Total cross sections of $p^{11}$Be radiative capture to the GS in the energy range 50 – 1.5 $10^3$ keV. Experimental data: closed triangles (▲) – [28], dots (●) – [27], open squares (□) – sum of cross sections [30], closed squares (■) – [29]. Curves correspond to the calculation of the total cross sections for the transitions to the GS with potentials given in text.

Corresponding results for the astrophysical *S*-factor are presented in Fig. 4. Experimental data for *S*-factor have been recalculated from the total cross sections with using of exact mass values of point-like particles. Let us remark, that below 100 keV *S*-factor is practically constant and has a value 3.5(1) keV b in average. Situation with the resonant behavior of astrophysical factor is completely analogous what corresponding cross sections reveal. It should be noted that it is not feasible yet to obtain a potential with 0.02 keV width. At a time, it is reasonable to encourage experimentalists for new measurements of this channel basing on the modern methods, as the majority of available data refer to 50-60ies.

**Conclusions**

It was demonstrated that quite transparent assumptions on a way of construction of $p^{11}$B interaction potentials with forbidden states [3] make it possible to reproduce reasonably well available experimental data [27-30] for the radiative capture to the GS of $^{12}$C in the energy region from 80 meV up to 1500 keV, as well as claim that the energy dependence of the observed cross sections occurs due to $^3S_1 \to {}^3P_0$ and $^3P_2 \to {}^3P_0$ transitions. It should be noted, that there are no experimental data on the asymptotic constants for the GS, so our estimations might be regarded as a preliminary proposal.

Thereby, MPCM based on the deep attractive potentials with forbidden states and coordinated with the spectrum of resonance levels mares it possible to convey properly the behavior of the experimental cross sections of $p^{11}$B radiative capture onto the GS of $^{12}$C in rather wide energy range. Furthermore, total cross sections have been described at resonance energies as a whole. Potentials for the ground state and first excited are in conformity with basic characteristics of $^{12}$C in $p^{11}$B channel, namely binding energy, charge radius and asymptotic constants.

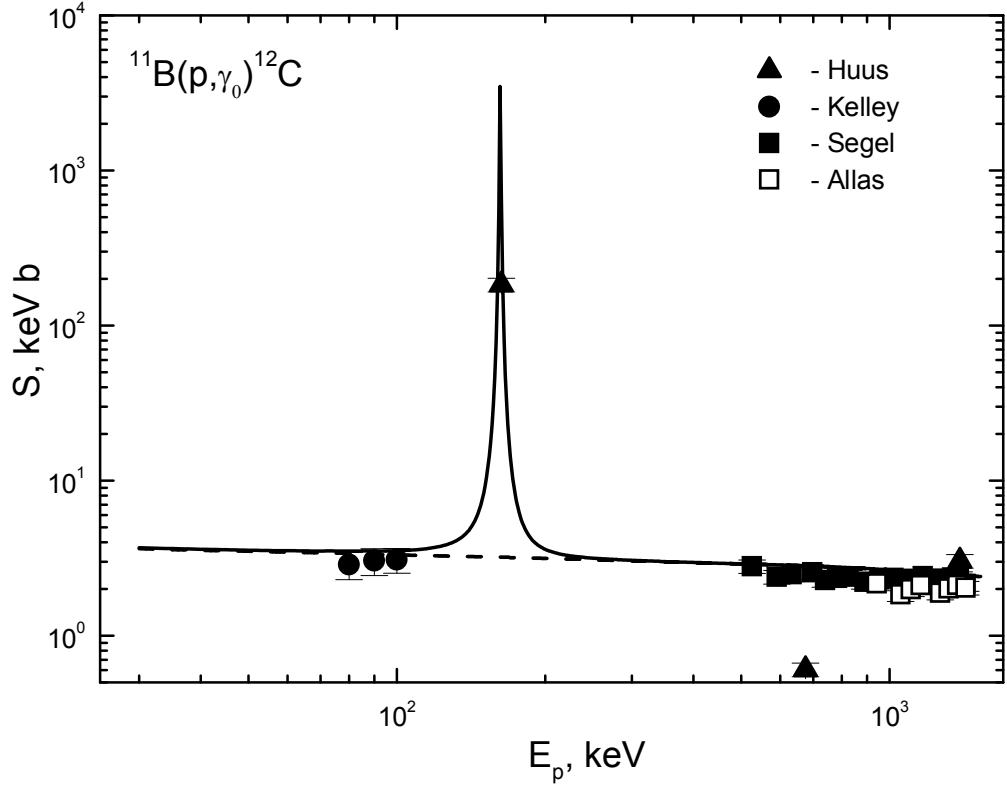

Fig. 4. Astrophysical *S*-factor of $p^{11}$Be radiative capture to the GS in the energy range 50 – 1.5 $10^3$ keV. Experimental data: closed triangles (▲) – [28], dots (●) – [27], open squares (□) – sum of cross sections [30], closed squares (■) – [29]. Curves correspond to the calculation of the total cross sections for the transitions to the GS with potentials given in text.

**Acknowledgements**

Authors express their heartfelt gratitude to Professor L.D. Blokhintsev and R. Yarmukhamedov for the providing very useful information on the AC in $p^{11}$B channel.

This work was supported by the Grant Program No. 0047/GF3 of the Ministry of Education and Science of the Republic of Kazakhstan: The Study of Main and Some Additional Thermonuclear Processes in the CNO Cycle on the Sun and Stars.